\documentclass[prl,twocolumn,superscriptaddress, 10pt, aps]{revtex4-2}
\usepackage[top=2.3cm,bottom=2.3cm, left = 1.5cm, right = 1.5cm]{geometry}

\usepackage{graphicx}
\usepackage{bm}
\usepackage[separate-uncertainty = true,multi-part-units=single]{siunitx}
\usepackage{xspace}
\usepackage{amssymb,amsfonts,amsmath}
\usepackage{siunitx}
\usepackage{MnSymbol}
\usepackage{xcolor}
\usepackage{hyperref}
\hypersetup{
    colorlinks=true,
    linkcolor=blue,
    filecolor=magenta,      
    urlcolor=cyan,
    citecolor = magenta,
    pdftitle={Overleaf Example},
    pdfpagemode=FullScreen,
    }

\usepackage{pdfpages} 
\usepackage{pgffor} 

\makeatletter
\AtBeginDocument{\let\LS@rot\@undefined}
\makeatother

\def\supplementfilename{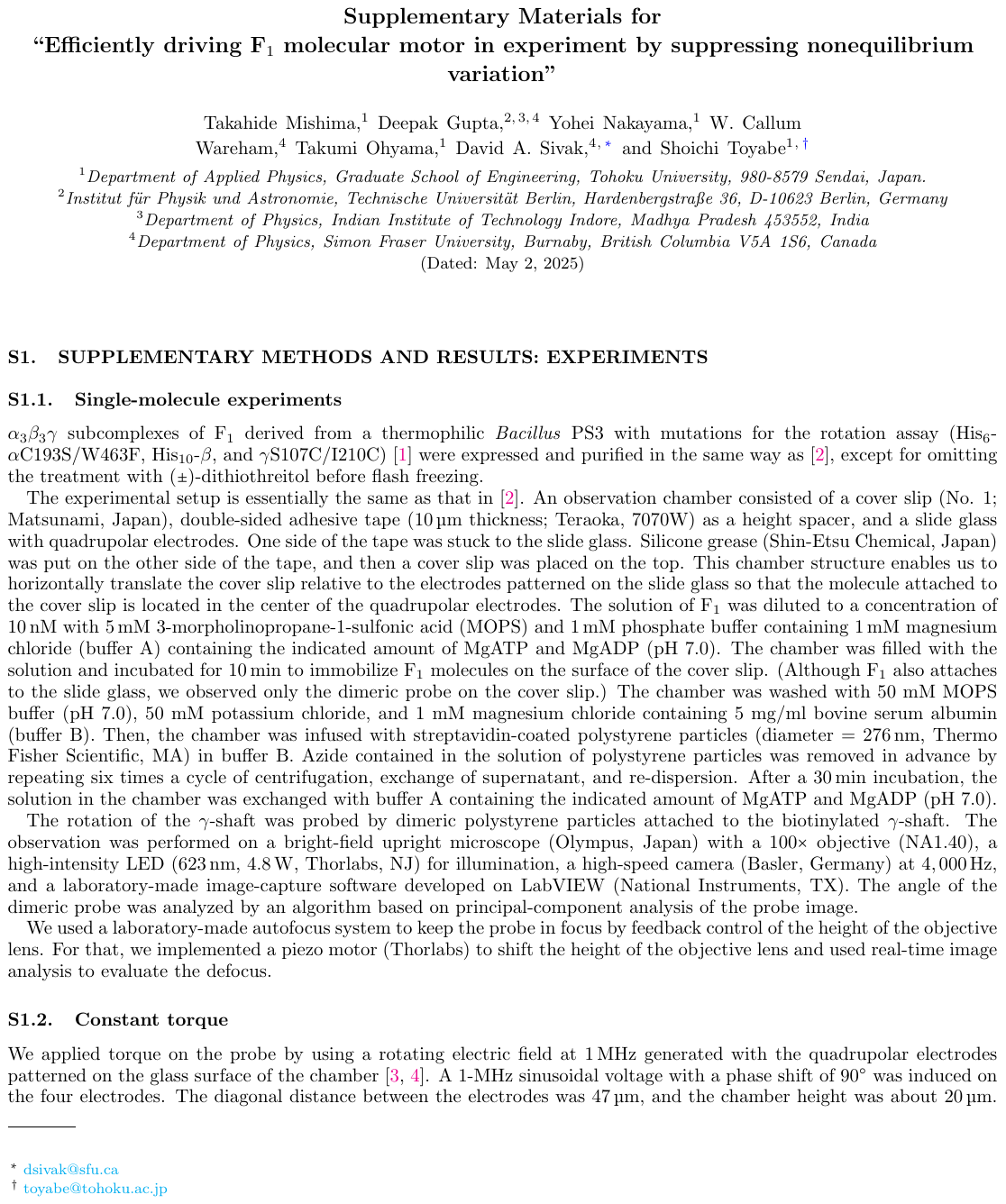}

\pdfximage{\supplementfilename}
\def\numbersupplementpages{\the\pdflastximagepages}

\newif\ifarXiv
\arXivtrue 

\usepackage[mathlines]{lineno}
\modulolinenumbers[10]
\newcommand*\patchAmsMathEnvironmentForLineno[1]{
  \expandafter\let\csname old#1\expandafter\endcsname\csname #1\endcsname
  \expandafter\let\csname oldend#1\expandafter\endcsname\csname end#1\endcsname
  \renewenvironment{#1}
     {\linenomath\csname old#1\endcsname}
     {\csname oldend#1\endcsname\endlinenomath}}
\newcommand*\patchBothAmsMathEnvironmentsForLineno[1]{
  \patchAmsMathEnvironmentForLineno{#1}
  \patchAmsMathEnvironmentForLineno{#1*}}
\AtBeginDocument{
\patchBothAmsMathEnvironmentsForLineno{equation}
\patchBothAmsMathEnvironmentsForLineno{align}
\patchBothAmsMathEnvironmentsForLineno{flalign}
\patchBothAmsMathEnvironmentsForLineno{alignat}
\patchBothAmsMathEnvironmentsForLineno{gather}
\patchBothAmsMathEnvironmentsForLineno{multline}
}

\begin{document}

\title{Efficiently driving F$_1$ molecular motor in experiment\\ by suppressing nonequilibrium variation}

\newcommand{\apph}{Department of Applied Physics, Graduate School of Engineering, Tohoku University, 980-8579 Sendai, Japan}
\newcommand {\addg}[1] {{\color {blue} #1}}

\author{Takahide Mishima}
\affiliation{\apph}
\author{Deepak Gupta}
\affiliation{Institut f\"ur Physik und Astronomie, Technische Universit\"at Berlin, Hardenbergstra\ss e 36, D-10623 Berlin, Germany}
\affiliation{Department of Physics, Indian Institute of Technology Indore, Madhya Pradesh 453552, India}
\affiliation{Department of Physics, Simon Fraser University, Burnaby, British Columbia V5A 1S6, Canada}
\author{Yohei Nakayama}
\affiliation{\apph}
\author{W. Callum Wareham}
\affiliation{Department of Physics, Simon Fraser University, Burnaby, British Columbia V5A 1S6, Canada}
\author{Takumi Ohyama}
\affiliation{\apph}
\author{David A.\ Sivak}
\email{dsivak@sfu.ca}
\affiliation{Department of Physics, Simon Fraser University, Burnaby, British Columbia V5A 1S6, Canada}
\author{Shoichi Toyabe}
\email{toyabe@tohoku.ac.jp}
\affiliation{\apph}

\date{\today}

\newcommand{\kB}{k_\mathrm{B}}
\newcommand{\kBT}{k_\mathrm{B}T}
\newcommand{\Wtorque}{W_\mathrm{trq}}
\newcommand{\Wangle}{W_\mathrm{ang}}
\newcommand{\Wdis}{W_\mathrm{d}}
\newcommand{\Pdis}{\dot W_\mathrm{d}}
\newcommand{\thetatrap}{\theta_\mathrm{trap}}
\newcommand{\vtrap}{v_\mathrm{trap}}
\renewcommand{\O}{\mathrm{o}}
\newcommand{\Fone}{F\textsubscript{1}\xspace}
\newcommand{\etaS}{\eta_\mathrm{Stokes}}
\newcommand{\Nnet}{N}
\newcommand{\NF}{N_\mathrm{F1}}
\newcommand{\Next}{N_\mathrm{ext}}
\newcommand{\Ndiff}{N_\mathrm{diff}}
\newcommand{\nudiff}{\nu_\mathrm{diff}}
\newcommand{\avg}[1]{\left\langle#1\right\rangle}
\newcommand{\aavg}[1]{\left\llangle#1\right\rrangle}
\newcommand{\Utrap}{U_\mathrm{trap}}
\newcommand{\dd}{\mathrm{d}}
\newcommand{\qq}{q}

\newcommand{\FigSAngleClamp}{S1}
\newcommand{\FigSBinwidth}{S2}
\newcommand{\FigSLocalMean}{S3}
\newcommand{\FigSFluctuation}{S4}
\newcommand{\FigSSimulation}{S5}

\begin{abstract}
F$_1$-ATPase (F$_1$) is central to cellular energy transduction.
Forcibly rotated by another motor F$_\O$, F$_1$ catalyzes ATP synthesis by converting mechanical work into chemical free energy stored in the molecule ATP.
The details of how F$_\O$ drives F$_1$ are not fully understood; however, evaluating efficient ways to rotate F$_1$ could provide fruitful insights into this driving since there is a selective pressure to improve efficiency.
Here, we show that rotating F$_1$ with an angle clamp is significantly more efficient than a constant torque.
Our experiments, combined with theory and simulation, indicate that the angle clamp significantly suppresses the nonequilibrium variation that contributes to the futile dissipation of input work.
\end{abstract}

\maketitle

\begin{figure}[!tbp]
\includegraphics{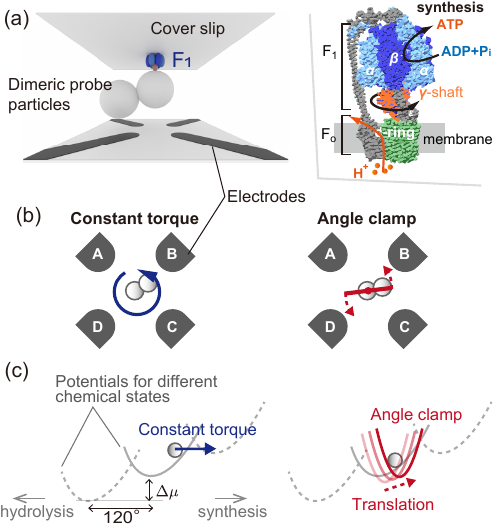}
\caption{Single-molecule control of F$_1$ as a model system of ATP synthase.
(a) We externally rotate F$_1$ derived from thermophilic {\it Bacillus} PS3~\cite{Noji_1997} by applying torque on the probe attached to its $\gamma$-shaft.
We observe the rotation by video microscopy at 4 kHz. Inset: F$_\mathrm{o}$F$_1$-ATP synthase.
(b) We realize constant-torque and angle-clamp driving modes by varying the amplitudes and phases of the alternating-current voltages induced on the four electrodes (A--D) \cite{Toyabe_2010}.
(c) The free-energy profile results from the interaction between the $\gamma$-shaft and stator of F$_1$ and the chemical free-energy change $\Delta\mu$ of ATP hydrolysis.
We rotate the probe attached to the $\gamma$-shaft in the ATP-synthetic direction by a constant torque (left) or by a trapping torque with the trap center rotated at a constant rate (right).}
\label{fig:intro}
\end{figure}

Nanomachines, such as biological molecular motors and artificially synthesized machines, are thought to utilize fluctuations to improve their performance in environments where fluctuations are dominant~\cite{Bustamante_2005}.
Hence, it is natural to think that efficient control of nanomachines differs from that of macroscopic engines.
The F$_\O$F$_1$-ATP synthase is a celebrated example of a biological nanomachine [Fig.~\ref{fig:intro}a, inset]. 
A prominent characteristic of this enzyme is that it consists of two mechanically coupled motors, F$_\O$ and F$_1$.
The c-ring of F$_\O$ rotates the connected $\gamma$-shaft in F$_1$, driven by an electrochemical gradient of H$^+$ ions across the membrane in which it is embedded.
F$_1$ then synthesizes adenosine triphosphate (ATP) molecules from adenosine diphosphate (ADP) and phosphate every 120$^\circ$ rotation of the $\gamma$-shaft by converting part of the induced mechanical work $W$ into chemical free energy $\Delta\mu$ embodied in an ATP molecule \cite{Rondelez_2005, Toyabe_2011, Saita_2015, Soga_2017}.

Even in such microscopic systems, the second law of thermodynamics still places fundamental limits on control \cite{Seifert_2012, Pigolotti_book, Oikawa2025}.
If F$_\O$ rotates the $\gamma$-shaft 120$^\circ$ quasistatically (infinitely slowly), the average work by F$_\O$ per $120^\circ$ rotation is $W=\Delta\mu$  \cite{Toyabe_2011, Saita_2015, Soga_2017}, but at finite speed $W>\Delta\mu$.
The mean excess work $\Wdis=W-\Delta\mu\ge 0$ results in entropy production and, therefore, quantifies the cost for driving the system out of equilibrium.
Importantly, $\Wdis$ depends on the history of how the $\gamma$-shaft is rotated, prompting interest in understanding optimal ways to rotate $\gamma$-shaft that minimize $\Wdis$ \cite{Gupta_2022, Wareham_2024}.

Although the structure of ATP synthase has been solved at high resolution~\cite{Hahn_2018}, the details of how F$_\O$ exerts a torque on F$_1$ remain to be understood.
Considering that ATP synthase is a primary energy converter in the cell, it would be natural to expect that evolution has improved its efficiency.
Therefore, exploring efficient control to rotate the $\gamma$-shaft should help us understand the mechanical coupling between F$_\O$ and F$_1$.

To quantify how driving affects the work, here we experimentally rotate the $\gamma$-shaft of isolated F$_1$ [SI Sec.S1~\cite{Note1}] in the ATP-synthetic direction as a model system of F$_\O$F$_1$ free-energy transduction and compare the work during two distinct driving modes
that represent extremes of control: \emph{constant-torque} mode dictates the driving strength, whereas \emph{angle-clamp} mode controls the mean rotation speed [Fig.~\ref{fig:intro}b].
The torque is induced on the $\gamma$-shaft through a dielectric probe attached to the shaft [Fig.~\ref{fig:intro}a]. With the relatively large size of the probe, the mechanochemical coupling can be effectively modeled so that ATP hydrolysis and synthesis respectively correspond to 120$^\circ$ shifts of the potential in opposite directions \cite{Toyabe_2012, Kawaguchi_2014} [Fig.~\ref{fig:intro}c, SI Sec.S2~\footnote{See Supplemental Material at [URL will be inserted by publisher] for supplementary methods and texts.}].

{\it Thermodynamics --} 
Constant-torque mode induces a constant torque, independent of the probe angle, via a rotating electric field \cite{Watanabe_Nakayama_2008, Toyabe_2010PRL, Toyabe_2011}.
In this case, the average rotation rate is not an independent variable but is determined by frictional drag and the net torque on the probe, which is the sum of the induced external torque and the torque exerted by F$_1$.
The work done on F$_1$ per 120$^\circ$-rotation by a constant torque $\Next$ is simply
\begin{align}\label{eq:W:torque}
    \Wtorque = \Next \cdot 120^\circ.
\end{align}

For the alternative angle-clamp mode, an electric field oscillating at high frequency induces a sinusoidal external potential $\Utrap(\theta-\thetatrap)=-\tfrac{1}{4}k \cos 2(\theta-\thetatrap)$ \cite{Toyabe_2010}. 
The potential has a period of 180$^\circ$ and hence two energetic minima at $\thetatrap$ and $\thetatrap+180^\circ$, corresponding to the orientation of the oscillating electric field.
Near minima, $\Utrap(\theta-\thetatrap)\simeq k(\theta-\thetatrap)^2/2$, so $k$ corresponds to the trap's spring constant.
We rotate $\thetatrap$ at a constant rate $\vtrap$, thereby dictating the probe's rotation rate (throughout, we exclude the rare trajectories that escape from the local potential well) \cite{Rondelez_2005, Saita_2015}.
The mean work per 120$^\circ$-rotation of the potential is 
\begin{align}\label{eq:W:angle}
    \Wangle= \bigg\langle\int_0^\tau\frac{\partial \Utrap(\theta-\thetatrap)}{\partial\thetatrap}\, \vtrap\, \mathrm{d}t\bigg\rangle.
\end{align}
Here, $\langle\cdot\rangle$ is an ensemble average over stochastic fluctuations of $\theta(t)$, and $\tau=120^\circ/v$ is the duration of the $120^\circ$ rotation, for mean angular velocity $v$.

\begin{figure}[tbp]
\includegraphics{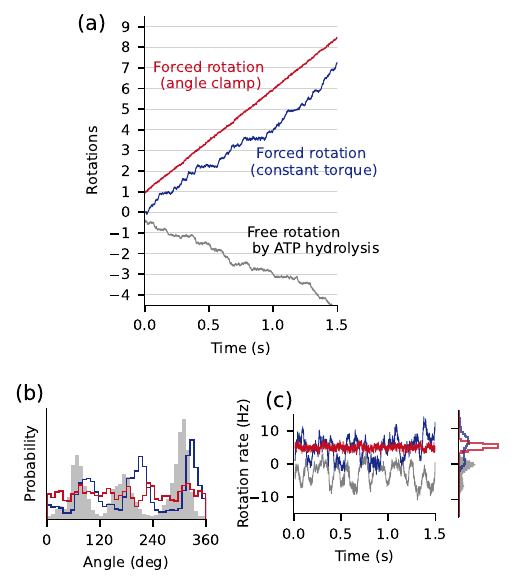}
\caption{Rotational trajectories and angular distributions. (a) Rotational trajectories, (b) angular distributions, and (c) rotational rates averaged over a \SI{0.05}{s} time window, for forced ATP-synthetic rotations under constant torque (blue) and angle clamp (red) and free ATP-hydrolytic rotations (gray).
[ATP] = [ADP] = \SI{0.4}{\micro M}.}
\label{fig:rotation}
\end{figure}

{\it Forced rotation --} 
The motor rotates in the ATP synthetic direction when subjected to a sufficiently strong constant torque or rotating angle clamp [Fig.~\ref{fig:rotation}a]; however, the rotational behavior differs significantly between the two modes.
On one hand, under constant torque, for low [ATP] and [ADP] F$_1$ takes three relatively discrete steps [Fig.~\ref{fig:rotation}b], as previously reported~\cite{Toyabe_2011}.
On the other hand, under the angle clamp the rotation is smooth, without distinct steps; accordingly, the velocity fluctuations are smaller [Fig.~\ref{fig:rotation}c].

\begin{figure}[tbp]
\includegraphics{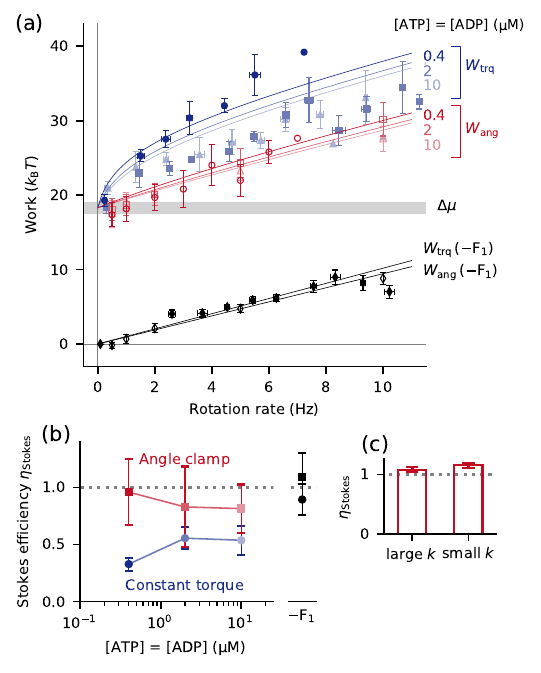}
\caption{(a) Average work to rotate the $\gamma$-shaft 120$^\circ$, for [ATP]=[ADP]=\SI{0.4}{\micro M} (circles), \SI{2}{\micro M} (squares), and \SI{10}{\micro M} (triangles) by constant torque (blue filled symbols) and angle clamp (red open symbols).
Solid curves show fits of simulations [SI Sec.~S2~\cite{Note1}].
The negative control $-\mathrm{F}_1$ corresponds to a freely rotating Brownian dimeric probe pinned on a glass surface without F$_1$.
159 trajectories (32 molecules) and 278 trajectories (34 molecules) are measured for the angle-clamp and constant-torque modes, respectively.
$\Wtorque$ is averaged across bins with 1-Hz width.
(b) The Stokes efficiency $\etaS$ [Eq.~\eqref{eq:eta:Stokes}] for different [ATP]=[ADP], calculated using the mean of $\Wdis$ for rotational rates between \SI{4}{Hz} and \SI{6}{Hz}.
(c) $\etaS$ in angle clamp was separately evaluated (using linear fit slope of work vs.\ global mean velocity, SI Sec.~S1.3~\cite{Note1}) for molecules with large or small $k$ values (threshold is $70\,\,\kBT/\mathrm{rad^2}$, and different [ATP]=[ADP] concentrations are mixed).
Error bars indicate standard errors.}
\label{fig:Work}
\end{figure}

{\it Work --} We measure $\Wtorque$ [Eq.~\eqref{eq:W:torque}] and $\Wangle$ [Eq.~\eqref{eq:W:angle}] under different rotation rates and [ATP]=[ADP] concentrations [Fig.~\ref{fig:Work}a].
Phosphate concentration $\mathrm{[P_i]=\SI{1}{mM}}$ is fixed, so $\Delta\mu=\Delta\mu^\circ+\kBT\ln(\mathrm{[ATP]/[ADP][P_i]})$ remains fixed at $\approx 18.3\kBT$,
for Boltzmann constant $\kB$ and absolute temperature $T$.
ATP and ADP are mixed in the forms of MgATP and MgADP, hence the concentrations of these chemical species influence the ionic strength, but this does not significantly affect the standard free energy $\Delta\mu^\circ$.
Both $\Wtorque$ and $\Wangle$ increase with the average rotation rate [Fig.~\ref{fig:rotation}a].
Under all tested conditions, $\Wtorque>\Wangle$ for a given rotation rate, with the difference higher for lower [ATP] and [ADP].

The Stokes efficiency 
\begin{align}\label{eq:eta:Stokes}
    \etaS \equiv \frac{\Gamma v\cdot 120^\circ}{\Wdis}
\end{align}
quantifies the efficiency of driving against viscous drag \cite{Wang_2002, Li2020}.
Here, $\Gamma$ is the rotational friction coefficient, $\Gamma v\cdot 120^\circ$ is the output work against viscous drag per $120^\circ$ rotation if the probe realized a constant angular velocity $v$, and $\Wdis$ is the input work consumed per $120^\circ$ rotation.
$\etaS=1$, in general, implies negligible angular variation of local mean velocity~\cite{Wang_2002} (see \emph{Nonequilibrium variation}).
$\etaS$ is evaluated by calculating the mean of $\Wdis$ across rotation frequencies between \SI{4}{Hz} and \SI{6}{Hz}.
The angle clamp has larger $\etaS$ than the constant torque [Fig.~\ref{fig:Work}b] and reaches a value similar to one, reflecting the small velocity variation [Fig.~\ref{fig:rotation}c].
Under constant torque, greater [ATP]=[ADP] concentrations give larger $\etaS$: at low [ATP] = [ADP], stepwise trajectories are observed, where in the dwell between steps the molecule awaits substrate binding, whereas at high [ATP] = [ADP], such dwells are short, and the trajectories are smoother.
Under the angle clamp, $\etaS$ does not significantly depend on substrate concentrations.

To understand these results, we simulated forced rotation using an established model \cite{Kawaguchi_2014} parametrized from single-molecule experimental data [Fig.~\FigSSimulation~\cite{Note1}].
The simulation qualitatively reproduces the experimental work measurements, validating the model [Fig.~\ref{fig:Work}a].

{\it Dependence on the trap stiffness --}
The spring constant $k$ of the angle clamp varies from molecule to molecule because $k$ depends on multiple factors, including the geometry of the dimeric probe.
We evaluated $\etaS$ in angle clamp separately for molecules with large or small $k$ values [Fig.~\ref{fig:Work}c].
To eliminate $k$-dependent calibration errors in $\thetatrap$ [SI Sec.~S1.3~\cite{Note1}], $\etaS$ was evaluated from the linear-fit slopes $\alpha$ of $W$ vs.\ $v$ as $\etaS=\Gamma\cdot 120^\circ/\alpha$.
$\etaS$ is not seen to significantly depend on $k$.
This insensitivity to $k$ is consistent with our simulations [Fig.~{\FigSSimulation}c~\cite{Note1}], which found only a slight decrease in $W$ (i.e., an increase in $\etaS$) with $k$.
We cannot access a broad range of $k$ values in both experiments and simulations because, for weak $k$, the probe easily slips out of the local minimum of $\Utrap(\theta)$; 
however, for harmonic $\Utrap(\theta)$ with small $k$, the external torque is only weakly dependent on $\theta$. 
Hence, an angle clamp with small $k$ may be similar to a constant torque, and it is expected that $\Wangle$ is larger for smaller $k$.

{\it Nonequilibrium variation --}
To understand the mechanism behind the angle clamp's advantage, we consider a Brownian particle on a one-dimensional potential for simplicity.
We define the local mean angular velocity $\nu(\theta, t)$ as the ensemble average of the angular velocity $\dd\theta(t)/\dd t$ at angle $\theta$ and time $t$ [SI Sec.~S1.4, Fig.~\FigSLocalMean~\cite{Note1}].
Constant torque produces a steady state, leading to $\nu(\theta, t)=\nu(\theta)$ independent of $t$.
In the angle-clamp mode, $\nu(\theta, t)$ is periodic in $t$ with a period of $3\tau$ (the duration of a 360$^\circ$-rotation of the angle clamp).

In this setup, $\Pdis$ coincides with the mean heat rate determined by the squared average of $\nu(\theta, t)$ \cite{Seifert_2012}:
\begin{align}\label{eq:Wdis:nu:1}
    \dot\qq\equiv \Gamma\aavg{\overline{\nu(\theta,t)^2}}.
\end{align}
Throughout, the dot on variables indicates a time-averaged rate.
For an arbitrary quantity $A(\theta, t)$, $A(\theta) = \overline {A(\theta, t)}\equiv \frac 1{3\tau p(\theta)}\int^{3\tau}_0 \mathrm{d}t \, A(\theta, t)p(\theta, t)$ is the local mean at $\theta$ over a full rotation of the angle clamp. 
The probability density $p(\theta, t)$ over $\theta$ at given $t$ has normalization $\int^{360^\circ}_{0^\circ} \mathrm{d}\theta \, p(\theta, t)=1$, and the time-averaged probability density over $\theta$ is $p(\theta)\equiv\frac 1{3\tau}\int^{3\tau}_0\mathrm{d}t \, p(\theta, t)$.
$\aavg{A(\theta)}\equiv\int^{360^\circ}_{0^\circ} \mathrm{d}\theta \, A(\theta)p(\theta)$ is the average over $\theta$ of an arbitrary quantity $A(\theta)$.
The local mean net torque exerted on the probe is $\Nnet(\theta, t)=\Gamma\nu(\theta, t)$, the sum of the torque by external driving ($\Next$) and F$_1$ ($\NF$), and the diffusional driving by the gradient of the probability density ($\Gamma\nudiff$). 
Here, $\nudiff=-\frac{\kBT}\Gamma\partial\ln p(\theta, t)/\partial \theta$ \cite{Seifert_2012}.
 
The decomposition $\aavg{\overline{\nu(\theta,t)^2}}=v^2+\aavg{[\nu(\theta)-v]^2}+\aavg{\overline{[\nu(\theta,t)-\nu(\theta)]^2}}$ in terms of local mean velocity, where $\nu(\theta)\equiv\overline{\nu(\theta, t)}$ and $v\equiv\aavg{\nu(\theta)}$, produces a decomposition in terms of local mean net torque:
\begin{align}\label{eq:Wdis:nu:2}
\begin{split}
\dot\qq=\underbrace{N_0^2/\Gamma}_{\dot\qq_0} +\underbrace{\aavg{[\Nnet(\theta)-N_0]^2}/\Gamma}_{\dot\qq_\theta}
+\underbrace{\aavg{\overline{[\Nnet(\theta,t)-\Nnet(\theta)]^2}}/\Gamma}_{\dot\qq_t}.
\end{split}
\end{align}
Here, $N(\theta)=\overline{N(\theta, t)}$ is the local mean torque at $\theta$, and $N_0=\aavg{N(\theta)}=\Gamma v$ is the global mean torque. 
The right-hand side terms respectively quantify the dissipation originating from global mean angular velocity $v$, angular variation of $\nu(\theta)$ around $v$, and temporal variation of $\nu(\theta,t)$ around $\nu(\theta)$ (respectively denoted as $\dot\qq_0$, $\dot\qq_\theta$, and $\dot\qq_t$). 

\begin{figure}[tbp]
{\centering\includegraphics{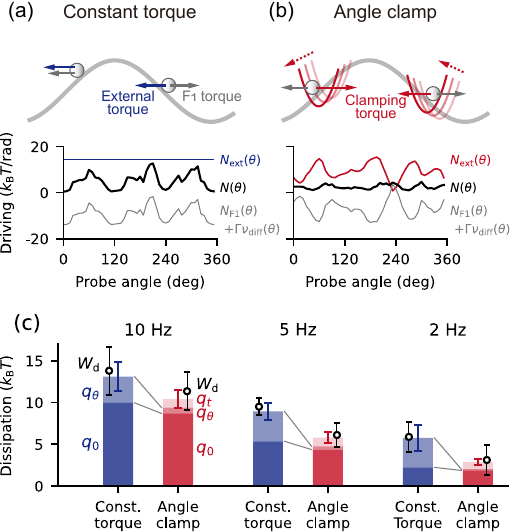}}
\caption{The mechanism for reduced work under angle clamp.
Schematic of the mechanism (top) and representative examples of torque balance (bottom) under constant torque (a) and angle clamp (b), for average rotation rate $\approx\SI{5}{Hz}$.
The net torque is obtained from the local mean velocity as $\Nnet(\theta)=\Gamma\nu(\theta)$.
Subtracting the external torque $\Next(\theta)$ gives the composite torque by F$_1$ and diffusional flow: $\NF(\theta)+\Ndiff(\theta)=\Nnet(\theta) - \Next(\theta)$.
(c) Dissipation estimates $\qq$ [Eq.~\eqref{eq:Wdis:nu:2}] (bars) and $\Wdis=W-\Delta\mu$ with $W$ calculated by Eq.~\eqref{eq:W:angle} (circles) per 120$^\circ$ rotation, for indicated average rotation rate and $\mathrm{[ATP]=[ADP]}=\SI{10}{\micro M}$.
$\qq$ and its decompositions are obtained by integrating the terms in Eq.~\eqref{eq:Wdis:nu:2} over 120$^\circ$ rotation.}
\label{fig:nettorque}
\end{figure}

Under constant-torque and angle-clamp modes, we independently evaluated $\Nnet(\theta)$ and $\Next(\theta)$, thereby quantifying the composite internal driving $\NF(\theta)+\Gamma\nudiff(\theta)=\Nnet(\theta)-\Next(\theta)$ [Fig.~\ref{fig:nettorque}a,b].
The angular variation of $\Nnet(\theta)$ under the angle clamp is significantly less than that under constant torque.
This can be intuitively understood: under constant (externally applied) torque, the net torque and hence the rotation rate is lower when facing resistive torque from F$_1$ and higher when receiving assistive torque (and these systematically vary across angles); however, under the angle clamp, the external trapping potential increases torque when the $\gamma$-shaft lags (at the ensemble level, due to resistive torque from F$_1$) and decreases torque when the $\gamma$-shaft rotates particularly rapidly (generally when F$_1$ provides assistive torque), thus acting as a suspension to suppress variation of the rotation rate.
By contrast, constant-torque driving acts without any `shocks'.

Accordingly, the angle clamp has significantly smaller dissipation $\qq_\theta$ due to angular variation [Fig.~\ref{fig:nettorque}c].
By contrast, dissipation $\qq_t$ due to temporal variation vanishes under constant torque since the ensemble is at steady state: $\Nnet(\theta, t)=\Nnet(\theta)$.
At high substrate concentrations such as \SI{10}{\micro M} (larger than the Michaelis constant $K_\mathrm{M} \approx$\SI{2}{\micro M} under experimental conditions~\cite{nakayama2024}), substrate binding does not significantly slow the chemical rates, and 
F$_1$'s mechanical dynamics is well-approximated using a one-dimensional potential.
Consistently, in this regime $\qq$ is similar to $\Wdis$ [Fig.~\ref{fig:nettorque}c].
While the angle clamp has significant temporal variation $\qq_t$, the total 
variation $\qq_\theta+\qq_t$ was greater for constant torque than for angle clamp.
The results demonstrate that the angle clamp reduces nonequilibrium variation by a suspension mechanism and, therefore, reduces the dissipation.

At high substrate concentrations, chemical reactions are sufficiently fast compared to rotational diffusion that dissipation by chemical switching is negligible. Then, $\nu(\theta, t)$ encompasses all the dissipation by Eq.~\eqref{eq:Wdis:nu:1}. 
At low substrate concentrations, switching between distinct mechanical potentials has an important effect.
Indeed, for 0.4 and $\SI{2}{\micro M}$ we find a significant difference between $\Wdis$ and apparent heat $\qq$, due to the unobserved chemical degree of freedom~\cite{Mehl2012}  [Fig.~\FigSFluctuation~\cite{Note1}].

Many attempts have been made to relate the magnitude of the nonequilibrium fluctuations to work or entropy production \cite{Seifert_2019}, based on, e.g., fluctuation theorems \cite{Crooks_1999}, the fluctuation-dissipation theorem \cite{Harada_2005}, or thermodynamic uncertainty relations (TURs) \cite{Barato_2015, Horowitz_2020}; however, it is not straightforward to apply these relations to the present setup.
For instance, the TUR provides an upper bound (determined by the diffusion coefficient) on the Stokes efficiency, but the TUR is generally valid only in steady state, and its application to angle-clamping mode is not straightforward.

{\it Discussion --}
In this paper, we explored efficient control of F$_1$ using single-molecule experiments and simulations.
We demonstrated that angle-clamp driving requires significantly less work than constant-torque driving. 
Further analysis revealed the crucial importance of suppressing variation in achieving efficient control. 

We rotated the angle clamp at a constant rate.
However, optimal control theory~\cite{Schmiedl_2007,Blaber_2023} indicates
that the work could be further reduced by modulating the clamp's instantaneous rotation rate and stiffness depending on F$_1$'s potential profile.
Guided by work-minimizing protocols in the linear-response regime~\cite{Sivak_2012}, uniform excess power over $t$ has been suggested as a general design principle for the optimal protocol.
This framework has been applied to DNA-pulling experiments~\cite{Tafoya_2019} and simulations of F$_1$~\cite{Gupta_2022, Wareham_2024} and active particles~\cite{OC_AP}. 
The forms of Eqs.~\eqref{eq:Wdis:nu:1} and \eqref{eq:Wdis:nu:2} suggest that uniform entropy production (instead of uniform excess power) is a natural guiding principle for minimizing work; the linear-response limit is a special case where all excess work is immediately dissipated as heat.
In contrast, constant torque drives the system to a nonequilibrium steady state.
In this case, work is minimized when power $\Next\nu(\theta)$ is uniform over $\theta$, given that the dynamics are approximated by a one-dimensional Langevin equation. 
Future studies should more broadly test the validity of these guiding principles.

The torque generated by F$_\O$ using the electrochemical gradient of H$^+$ ions is transmitted to the $\gamma$-shaft of F$_1$ via an elastic coupling; this coupling naturally suggests an angle-clamp driving mechanism. 
The smaller work by angle clamp also suggests a selective advantage for this driving mechanism.
However, F$_\O$ (itself a strongly fluctuating object) can only implement stochastic driving~\cite{large_stochastic_2018}, and in the intact F$_\O$F$_1$ holoenzyme F$_\O$'s driving is 
not independent of F$_1$.
Thus, the true \emph{in vivo} situation is likely more complicated than our angle-clamp experiments.
Future experiments, including the simultaneous observation of the rotations of F$_\O$ and F$_1$, should help clarify the intervening torque-transmission mechanism.

We thank Alex Tong and Carlos Bustamante for helping motivate this project. This work was supported by JSPS KAKENHI Grant Numbers JP18H05427, JP23K25833, and JST ERATO Grant Number JPMJER2302 (all to S.T.); the Alexander von Humboldt Foundation (D.G.); Natural Sciences and Engineering Research Council of Canada (NSERC) CGS Master's and Doctoral Fellowships (W.C.W.); and an NSERC Alliance International Collaboration Grant ALLRP-2023-585940, an NSERC Discovery Grant and Discovery Accelerator Supplement RGPIN-2020-04950, and a Tier-II Canada Research Chair CRC-2020-00098 (all to D.A.S.); JSPS KAKENHI Grant Numbers JP24K06971 (to Y.N.). This research was enabled in part by support provided by BC DRI Group and the Digital Research Alliance of Canada (www.alliancecan.ca).

\nocite{Aurell2012}

\bibliographystyle{prsty}
\bibliography{main}

\ifarXiv
    

\section*{Supplementary material}

# Supplementary Materials for "Efficiently driving $F_1$ molecular motor in experiment by suppressing nonequilibrium variation"

Takahide Mishima,[1] Deepak Gupta,[2, 3, 4] Yohei Nakayama,[1] W. Callum Wareham,[4] Takumi Ohyama,[1] David A. Sivak,[4, *] and Shoichi Toyabe[1, †]

[1]*Department of Applied Physics, Graduate School of Engineering, Tohoku University, 980-8579 Sendai, Japan.*
[2]*Institut für Physik und Astronomie, Technische Universität Berlin, Hardenbergstraße 36, D-10623 Berlin, Germany*
[3]*Department of Physics, Indian Institute of Technology Indore, Madhya Pradesh 453552, India*
[4]*Department of Physics, Simon Fraser University, Burnaby, British Columbia V5A 1S6, Canada*

(Dated: May 2, 2025)

## S1. SUPPLEMENTARY METHODS AND RESULTS: EXPERIMENTS

### S1.1. Single-molecule experiments

$\alpha_3\beta_3\gamma$ subcomplexes of $F_1$ derived from a thermophilic *Bacillus* PS3 with mutations for the rotation assay (His$_6$-$\alpha$C193S/W463F, His$_{10}$-$\beta$, and $\gamma$S107C/I210C) [1] were expressed and purified in the same way as [2], except for omitting the treatment with (±)-dithiothreitol before flash freezing.

The experimental setup is essentially the same as that in [2]. An observation chamber consisted of a cover slip (No. 1; Matsunami, Japan), double-sided adhesive tape (10 µm thickness; Teraoka, 7070W) as a height spacer, and a slide glass with quadrupolar electrodes. One side of the tape was stuck to the slide glass. Silicone grease (Shin-Etsu Chemical, Japan) was put on the other side of the tape, and then a cover slip was placed on the top. This chamber structure enables us to horizontally translate the cover slip relative to the electrodes patterned on the slide glass so that the molecule attached to the cover slip is located in the center of the quadrupolar electrodes. The solution of $F_1$ was diluted to a concentration of 10 nM with 5 mM 3-morpholinopropane-1-sulfonic acid (MOPS) and 1 mM phosphate buffer containing 1 mM magnesium chloride (buffer A) containing the indicated amount of MgATP and MgADP (pH 7.0). The chamber was filled with the solution and incubated for 10 min to immobilize $F_1$ molecules on the surface of the cover slip. (Although $F_1$ also attaches to the slide glass, we observed only the dimeric probe on the cover slip.) The chamber was washed with 50 mM MOPS buffer (pH 7.0), 50 mM potassium chloride, and 1 mM magnesium chloride containing 5 mg/ml bovine serum albumin (buffer B). Then, the chamber was infused with streptavidin-coated polystyrene particles (diameter = 276 nm, Thermo Fisher Scientific, MA) in buffer B. Azide contained in the solution of polystyrene particles was removed in advance by repeating six times a cycle of centrifugation, exchange of supernatant, and re-dispersion. After a 30 min incubation, the solution in the chamber was exchanged with buffer A containing the indicated amount of MgATP and MgADP (pH 7.0).

The rotation of the $\gamma$-shaft was probed by dimeric polystyrene particles attached to the biotinylated $\gamma$-shaft. The observation was performed on a bright-field upright microscope (Olympus, Japan) with a 100× objective (NA1.40), a high-intensity LED (623 nm, 4.8 W, Thorlabs, NJ) for illumination, a high-speed camera (Basler, Germany) at 4,000 Hz, and a laboratory-made image-capture software developed on LabVIEW (National Instruments, TX). The angle of the dimeric probe was analyzed by an algorithm based on principal-component analysis of the probe image.

We used a laboratory-made autofocus system to keep the probe in focus by feedback control of the height of the objective lens. For that, we implemented a piezo motor (Thorlabs) to shift the height of the objective lens and used real-time image analysis to evaluate the defocus.

### S1.2. Constant torque

We applied torque on the probe by using a rotating electric field at 1 MHz generated with the quadrupolar electrodes patterned on the glass surface of the chamber [3, 4]. A 1-MHz sinusoidal voltage with a phase shift of 90° was induced on the four electrodes. The diagonal distance between the electrodes was 47 µm, and the chamber height was about 20 µm.

---

[*] dsivak@sfu.ca
[†] toyabe@tohoku.ac.jp

The signals generated by a function generator (WF1974; nf, Japan) were divided by a 180° phase divider (Thamway, Japan), amplified by four bipolar amplifiers (BA4825; nf, Japan), and loaded on the electrodes. This generates an electric field rotating at 1 MHz in the center of the electrodes and induces a dipole moment rotating at 1 MHz on the dimeric probe. Since there is a phase delay of the dipole moment with respect to the electric field, the dimeric probe is subjected to a constant torque. The torque magnitude was regulated by a multifunction board (National Instruments, TX) equipped on a computer, which controls the voltage amplitude $V_0$ of the signals. The torque magnitude is proportional to the square of $V_0$ [3, 4]. The camera and amplitude signal were synchronized at a time difference of less than 1 µs. We calibrated the torque magnitude based on the fluctuation-response relation at 500 Hz as previously reported [4]; the response to the 500-Hz external probe torque with a sinusoidal temporal profile is compared to the fluctuation around 500 Hz. Since equilibration is expected in such high-frequency regimes, the fluctuation and response are related by the fluctuation-response relation [5] and give the proportionality coefficient between $V_0^2$ and the corresponding torque magnitude. We estimated the friction coefficient $\Gamma$ using the same data because the amplitude of the angular-velocity fluctuation spectrum is $2k_\mathrm{B}T/\Gamma$ (the coefficient depends on the definition of the frequency space) in the high-frequency region. We obtained $\Gamma = 0.076 \pm 0.011\ k_\mathrm{B}T \cdot \mathrm{s}/\mathrm{rad}^2$ (mean ± SD, $N$ = 278 trajectories (34 molecules)).

### S1.3. Angle clamp

We implemented an angle clamp using the same setup as for constant torque, but chose the phases of the voltages on the electrodes so that the electric field oscillates at a specific angle at 3 MHz. The voltages of the electrodes facing each other are chosen to be in phase opposition, i.e., 180° out of phase and at the same amplitude. To change the clamp angle, we change the relative voltage amplitudes of the two opposing pairs.

To calibrate the angle of the oscillating electric field, from images of the electrodes, we measured the electrode angle relative to the camera's axis. However, the axis probed by the image analysis and the axis of the dielectric characteristics can slightly differ since each particle of the dimeric particle complex is not perfectly spherical. In estimating work by Eq. (2), this calibration error of $\theta_\mathrm{trap}$ may cause an error that grows with $k$. However, the results that as $v \to 0$ the work $W$ converges to zero for -$F_1$ and to $\Delta\mu$ for +$F_1$ [Fig. 3a] show that this calibration error does not make a systematic error in evaluating $W$.

This method generates a sinusoidal potential with 180° period, with the two minima corresponding to the oscillation direction of the electric field [6]. We calibrated the potential depth for each molecule before and after the work measurement. For the calibration, we first measured the spring constant $k$ from the angular distribution at a fixed clamp angle with voltage amplitude $V_0$ [Fig. S1a, b]. $k$ is the sum of the angle-clamp's spring constant $aV_0^2$, where $a$ is a proportionality coefficient to be determined, and the spring constant $k_{F1}$ due to the $F_1$ molecule's potential:

$$k(V_0) = aV_0^2 + k_{F1}. \tag{S1}$$

We measured $k(V_0)$ for different (typically five) $V_0$ values and obtained $a$ by a linear fit of $k(V_0)$ [Fig. S1d].

We also use these trajectories to estimate the friction coefficient. Let $C(\omega)$ be the fluctuation spectrum of the probe angle $\theta(t)$ as a function of angular frequency $\omega$, which is obtained from the Fourier transform of $\theta(t)$ through the Wiener–Khinchin theorem. Here, we define the Fourier transform of an arbitrary function $A(t)$ as $A(\omega) = \int_{-\infty}^{\infty} A(t)e^{i\omega t}$. If we approximate the potential as a harmonic potential, $C(\omega)$ is modeled by a Lorentzian function:

$$C_\mathrm{L}(\omega) \equiv \frac{2k_\mathrm{B}T}{\Gamma} \cdot \frac{1}{\omega^2 + \omega_0^2}, \qquad \omega_0 \equiv \frac{k}{\Gamma}. \tag{S2}$$

To estimate $\Gamma$, we fitted experimental $\omega C(\omega)$ by a least-squares method in the frequency range between 20 Hz and 500 Hz by $\omega C_\mathrm{L}(\omega)$ with fitting prameters $\Gamma$ and $\omega_0$ [Fig. S1c]. The multiplication by $\omega$ makes a peak around $\omega_0$ in the spectrum [Fig. S1c], which emphasizes the importance of the spectrum around $\omega_0$, which is expected to make the residual in fitting more sensitive to the value of $\omega_0$ and thereby give better parameter estimation. We obtained $\Gamma = 0.077 \pm 0.016\ k_\mathrm{B}T \cdot \mathrm{s}/\mathrm{rad}^2$ (mean ± SD, $N$ = 159 trajectories (32 molecules)). The spectrum method gives also an estimate of the spring constant $k$ as $\Gamma\omega_0$. The values of $k$ obtained from the distribution [Fig. S1b] and the spectrum [Fig. S1c] were similar [Fig. S1d], suggesting the validity of these methods.

### S1.4. Estimation of dissipation from angular and temporal variation of local mean velocity

We evaluated $q$ from the local mean velocity $\nu(\theta, t)$ using Eq. (5).

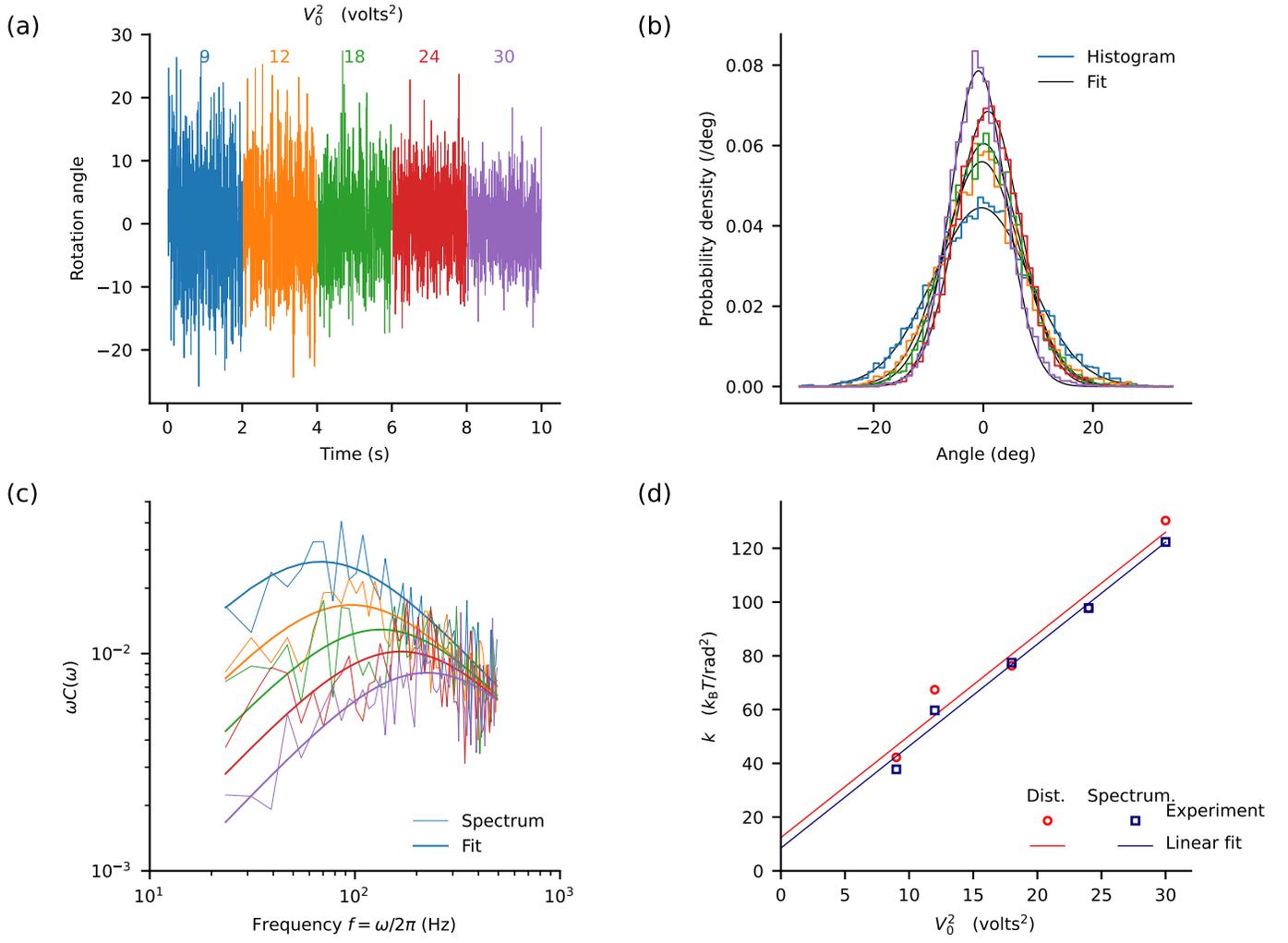

FIG. S1. Typical example of spring-constant calibration in angle clamp. (a) Rotational trajectories in angle clamp with different voltages. (b) Angular distributions. The thin curves are fits to $a\exp\{-\frac{1}{4}\beta k\cos 2(\theta - b)\}$ with fitting parameters $a$, $b$, and $k$. (c) Fluctuation spectrum of $\theta(t)$ multiplied by $\omega$, $\omega C(\omega)$. The curves are fits of $\omega C_{\mathrm{L}}(\omega)$ with the Lorentzian function $C_{\mathrm{L}}(\omega)$ [Eq. (S2)]. (d) Spring constant $k$ as a function of squared voltage $V_0^2$ (symbols) with linear fits (lines). (a), (b), and (c) share same color scheme.

The local mean velocity $\nu(\theta, t)$ is defined as [7]

$$\nu(\theta, t) \equiv \lim_{\epsilon \to 0} \frac{\langle [\theta(t+\epsilon) - \theta(t-\epsilon)]\delta(\theta(t) - \theta)\rangle}{2\epsilon \langle \delta(\theta(t) - \theta)\rangle}. \tag{S3}$$

Here, $\delta(\cdot)$ is the Dirac delta, and $\langle \cdot \rangle$ denotes an ensemble average over stochastic fluctuations of $\theta$. In the experiments, we evaluated $\nu(\theta, t)$ by approximating the right-hand side of Eq. (S3) as

$$\nu(\theta, t) \approx \frac{\left\langle \frac{\theta^{(n+1)} - \theta^{(n)}}{\Delta t} \delta\big(\frac{1}{2}[\theta^{(n+1)} + \theta^{(n)}] - \theta\big)\right\rangle}{\left\langle \delta\big(\frac{1}{2}[\theta^{(n+1)} + \theta^{(n)}] - \theta\big)\right\rangle}. \tag{S4}$$

Here, $\Delta t$ is the time interval between video frames, $\theta^{(n)}$ is the angle at the $n$th frame, and $n = t/\Delta t$. We divided $\theta$ and $t$ into bins and calculated the ensemble averages in Eq. (S4) in each $(\theta, t)$ bin. We also evaluated $p(\theta, t)$ by counting in each bin the number of samples $\theta(t)$ at time $t$.

To find the optimal bin width, we evaluated the dependence of dissipation $q$ on the bin width [Fig. S2]. On the one hand, $q$ increases steeply for smaller bin widths due to insufficient sampling in each bin. On the other hand, $q$ decreases as the bin width increases since temporal and angular variation of $\nu(\theta,t)$ is averaged out, decreasing $q_\theta$ and $q_t$. We chose a bin width of 7.5°, which falls at the boundary between the two regimes mentioned above. A slight change in the bin width does not qualitatively affect the conclusion shown in Fig. 4c that $W_\mathrm{d} \approx q$ at large [ATP]=[ADP]. Figure S3 shows a typical example of $\nu(\theta,t)$ and $p(\theta,t)$ for the angle clamp.

We observed significant differences between $q$ and $W_\mathrm{d}$ for 0.4 µM and 2 µM [ATP]=[ADP] [Fig. S4], suggesting that chemical switching of the potentials significantly contributes to the dissipation.

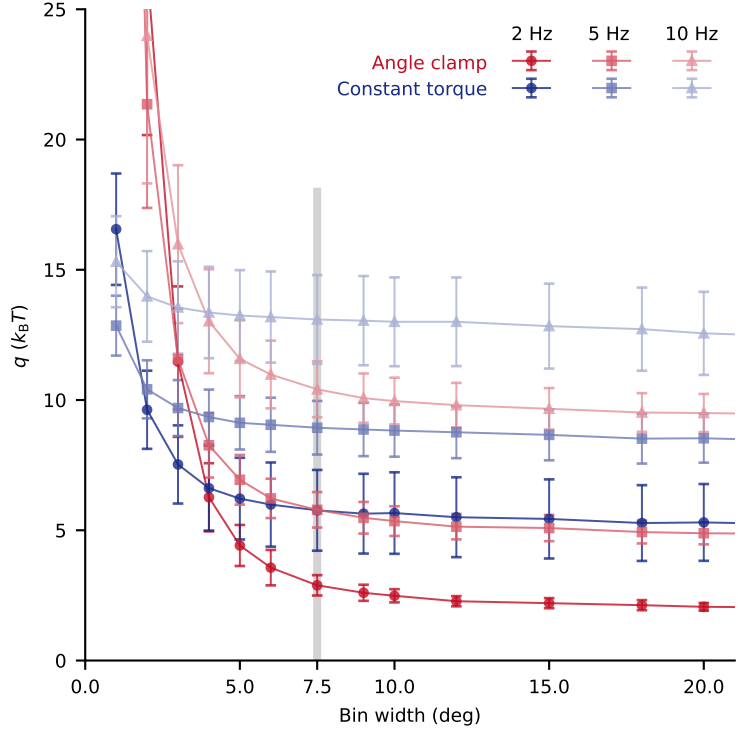

FIG. S2. Dissipation $q$ as a function of the bin width used for evaluating $\nu(\theta,t)$. [ATP] = [ADP] = 10 µM Under the angle clamp, the same value of the bin width is used for both $\theta_\mathrm{trap}$ and $\theta$. We used a bin width of 7.5° throughout the paper (indicated by a vertical gray line). Error bars indicate standard errors.

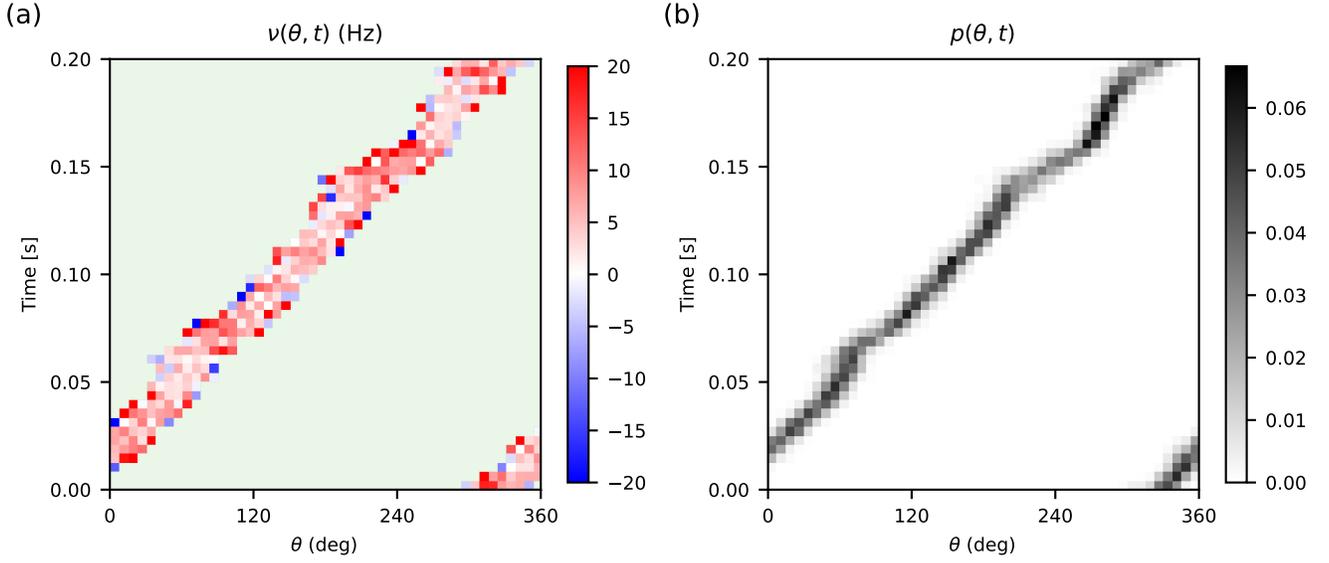

FIG. S3. A typical example of angle-clamp local mean velocity and probability distribution, at rotation rate 5 Hz (period = 0.2 s) and [ATP] = [ADP] = 10 μM. The bin width for the probe angle $\theta$ is 7.5°. The bin width for time is $\frac{7.5°}{360°} \cdot 3\tau = 0.0042$ s, which corresponds to a bin width of 7.5° for $\theta_{\text{trap}}$, since $\theta_{\text{trap}}$ rotates 360° in $3\tau$. (a) Local mean velocity at probe angle $\theta$ and time $t$. Light green: bins without samples, where $\nu(\theta, t)$ is not evaluated. (b) Probability distribution $p(\theta, t)$, normalized at each time $t$.

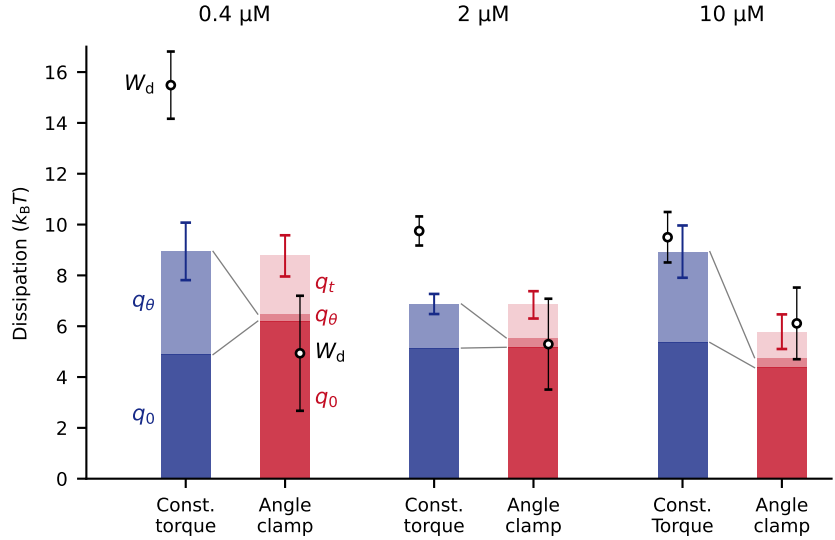

FIG. S4. Dissipation per 120° rotation, estimated as $q$ by Eq. (5) (bars) and $W_{\text{d}} \equiv W - \Delta\mu$ (circles), for indicated substrate concentrations ([ATP] = [ADP]) and $v = 5$ Hz. $W_{\text{d}}$ differs from $q$ except for 10 μM because the assumption of local equilibrium for the chemical reactions is only valid at large substrate concentrations. 10 μM bars duplicate Fig. 4c. The error bars indicate standard errors.

## S2. SUPPLEMENTARY METHODS AND RESULTS: SIMULATION

### S2.1. Computational model

The totally asymmetric allosteric model (TASAM) [8] of $F_1$ describes the evolution of a rotational degree of freedom $\theta$, corresponding to the probe attached to the $\gamma$-shaft. Following previous work [8], we model the full 360° rotation of the motor as three 120° steps between effective potentials that each coarse-grain over the 80° (slow) and 40° (fast) substeps. The effective potential is constructed by assuming equilibration between two harmonic potentials of the same spring constant $k_0 = 20\ k_\mathrm{B}T/\mathrm{rad}^2$ with minima separated by $\ell = 40°$ and offset by the free-energy difference $\widetilde{\Delta\mu} = 5.2\ k_\mathrm{B}T$:

$$\beta U_n(\theta) \equiv -\ln\left[e^{-\frac{1}{2}\beta k_0(\theta-\ell-n\xi)^2 - \beta\widetilde{\Delta\mu}} + e^{-\frac{1}{2}\beta k_0(\theta-n\xi)^2}\right]\ , \tag{S5}$$

for inverse temperature $\beta \equiv 1/(k_\mathrm{B}T)$ and $n = 0, \pm 1, \pm 2, \ldots$. The first and second terms in square brackets respectively represent the so-called catalytic-dwell and binding-dwell states. The coarse-grained 120° step is modeled by the system switching between adjacent effective potentials $U_n(\theta)$ and $U_{n\pm 1}(\theta)$ angularly separated by $\xi = 120°$ and offset by the chemical-potential difference $\Delta\mu \geq 0$.

Thus, the switching $U_n(\theta) \to U_{n\pm 1}(\theta)$ between effective potentials models the overall synthesis/hydrolysis of one ATP molecule, where the '+' ('−') sign denotes the synthesis (hydrolysis) mechanism. The corresponding transition rates obey the local detailed-balance condition:

$$\frac{R_n^+(\theta)}{R_{n+1}^-(\theta)} = e^{\beta[U_n(\theta)-U_{n+1}(\theta)-\Delta\mu]}\ , \tag{S6}$$

for synthesis transition rate

$$R_n^+(\theta) = w_0 e^{\beta[U_n(\theta)-U_{n+1}(\theta)-\Delta\mu]} \tag{S7}$$

from $U_n(\theta)$ to $U_{n+1}(\theta)$, and hydrolysis transition rate

$$R_{n+1}^-(\theta) = w_0 \tag{S8}$$

from $U_{n+1}(\theta)$ to $U_n(\theta)$. The rate $w_0$ characterizes the chemical reaction and is proportional to [ATP] (see Fig. S1 in Ref. [9]).

In our experiments, the dimeric probe attached to $F_1$'s $\gamma$-shaft is driven by two different techniques: 1) confining the probe with an *angle clamp* whose minimum is dynamically rotated, or 2) by applying a *constant torque* $N_\mathrm{ext}$ to the probe.

We model the angle clamp using a sinusoidal potential:

$$U_\mathrm{trap}(\theta - \theta_\mathrm{trap}) \equiv -\tfrac{1}{4}k\cos 2(\theta - \theta_\mathrm{trap})\ , \tag{S9}$$

for probe angle $\theta(t)$ and trap local minimum $\theta_\mathrm{trap}(t)$, which is a time-dependent control parameter. We change $\theta_\mathrm{trap}(t)$ at a constant rate $v_\mathrm{trap}$. The trap has two minima separated by 180° with barriers of height $k/2$ (parameterizing trap strength).

Subject to either driving mode, $\theta$ dynamically evolves according to

$$\Gamma\frac{\mathrm{d}\theta}{\mathrm{d}t} = -\partial_\theta U_n(\theta) + \sqrt{2k_\mathrm{B}T\Gamma}\,\eta(t) + \begin{cases}N_\mathrm{ext} & \text{constant torque,} \\ -\partial_\theta U_\mathrm{trap}(\theta - \theta_\mathrm{trap}) & \text{angle clamp,}\end{cases} \tag{S10}$$

for rotational friction constant $\Gamma = 0.073\ \mathrm{rad}^2/\mathrm{s}$ [10] and standard Gaussian white noise $\eta(t)$ with zero mean and unit variance. We only rotate the trap in the synthesis direction, which for constant torque requires sufficiently strong torque ($N_\mathrm{ext} \cdot 120° > \Delta\mu_\mathrm{ATP}$).

### S2.2. Potential of mean force: $w_0 \to \infty$

For high [ATP] (manifesting in the model as $w_0 \to \infty$), the chemical transitions occur rapidly (compared to mechanical rotation) and can be integrated out, producing effective rotational dynamics

$$\Gamma\frac{\mathrm{d}\theta}{\mathrm{d}t} = -\partial_\theta U_\mathrm{PMF}(\theta) + \sqrt{2k_\mathrm{B}T\Gamma}\,\eta(t) + \begin{cases}N_\mathrm{ext} & \text{constant torque ,} \\ -\partial_\theta U_\mathrm{trap}(\theta - \theta_\mathrm{trap}) & \text{angle clamp ,}\end{cases} \tag{S11}$$

for the potential of mean force (PMF)

$$\beta U_\mathrm{PMF}(\theta) \equiv -\ln\sum_{n=-\infty}^{+\infty} e^{-\beta[U_n(\theta)+n\Delta\mu_\mathrm{ATP}]}\ . \tag{S12}$$

### S2.3. Simulation methods

We discretize the Langevin equation (S10) to first order in $\Delta t$ by a forward Euler method:

$$\theta_m = \theta_{m-1} - \Gamma^{-1}\partial_\theta U_n(\theta_{m-1})\Delta t + \sqrt{\frac{2k_\mathrm{B}T}{\Gamma}\Delta t}\,\tilde{\eta}_{m-1} + \Gamma^{-1}\Delta t \begin{cases} N_\mathrm{ext} & \text{constant torque,} \\ -\partial_\theta U_\mathrm{trap}(\theta_{m-1}-\theta_{\mathrm{trap},m-1}) & \text{angle clamp,} \end{cases} \quad (S13)$$

Here the discrete simulation step is indexed by $m \in [1, 2, \ldots, \mathcal{T}/(\Delta t)]$ for trajectory duration $\mathcal{T}$, and $\tilde{\eta}_m$ is the normal random variable with zero mean, unit variance, and temporal correlation $\langle \tilde{\eta}_m \tilde{\eta}_{m'} \rangle = \delta_{m,m'}$ (for Kronekker delta $\delta_{m,m'}$. (For $[\text{ATP}] \to \infty$ $[w_0 \to \infty]$, we instead discretize Eq. (S11) in an analogous fashion.) The shaft angle is initialized to $\theta_0 = 0$. For each time increment $\Delta t$, we switch the potential $U_n \to U_{n\pm 1}$ according to probabilities $R_n^+(\theta_{m-1})\Delta t$ (S7) and $R_n^-(\theta_{m-1})\Delta t$ (S8). Then, we evolve one timestep $\Delta t$ of the discretized dynamics (S13). We repeat many such cycles and compute the angular velocity $d\theta/dt \equiv \theta_m/(m\Delta t)$ as a function of observation time. In here and what follows, $d\theta/dt > 0$ corresponds to ATP-synthetic rotation.

For the angle clamp, we then rotate the trap minimum according to

$$\theta_{\mathrm{trap},m} = \theta_{\mathrm{trap},m-1} + v_\mathrm{trap}\Delta t\,, \tag{S14}$$

for $\theta_{\mathrm{trap},0} = 0$ and a constant angular velocity $v_\mathrm{trap} > 0$ for the trap. To compute the external work due to the trap along a single stochastic trajectory, we compute the net energy difference arising due to change in the control parameter $\theta_\mathrm{trap}$ at fixed probe angle $\theta$:

$$w = \sum_{m=0}^{\mathcal{T}/(\Delta t)-1} \left[ U_\mathrm{trap}(\theta_{m+1} - \theta_{\mathrm{trap},m+1}) - U_\mathrm{trap}(\theta_{m+1} - \theta_{\mathrm{trap},m}) \right]. \tag{S15}$$

We average the rate of work, $\langle w \rangle / \mathcal{T}$, over 'surviving' trajectories that stay in the local potential well, i.e, for which $\theta_m \in [\theta_{\mathrm{trap},m-1} - 90°, \theta_{\mathrm{trap},m-1} + 90°]$. Finally, we compute the average work per 120° of the synthetic rotation by

$$W = \frac{\langle w \rangle / \mathcal{T}}{v} \cdot 120°, \tag{S16}$$

for global mean velocity $v$.

Under constant torque, the average external work on the probe per 120° rotation is

$$W = N_\mathrm{ext} \cdot 120°. \tag{S17}$$

For both constant torque and angle clamp, we record the average synthetic rate and the average work in the stationary state, i.e., when $v$ becomes independent of time.

We choose the discretization time step $\Delta t = 10^{-6}$ s. With this $\Delta t$, $w_0 \Delta t \ll 1$ even for the largest possible potential switching rate $w_0 = 10^3$ s$^{-1}$ from $U_n$ to $U_{n+1}$. We evolve the dynamics (S13) for $\mathcal{T} = 10$ s such that the system reaches the stationary state, i.e., $v$ becomes independent of time. We average over $10^3$ trajectories.

### S2.4. Simulation results and fitting to experiment

Figure S5a shows the simulation results as symbols.

We fitted experimental data with simulation curves. A fitting parameter is the proportional factor $s$ between $w_0$ and [ATP] (=[ADP]): $w_0 = s[\text{ATP}]$. We modeled the simulation curves by simple relations to find $s$ based on experimental curves. We first fitted the simulation curves with the equations

$$W(v) = p\ln[a(w_0)v + 1] + \Delta\mu + qv, \tag{S18a}$$
$$a(w_0) = bw_0^c + d, \tag{S18b}$$

where $p$, $q$, $b$, $c$, and $d$ are global fitting parameters common to different $w_0$ values ($p$ and $q$ are common for angle clamp and constant torque, but $b$, $c$ and $d$ are not). $W$ is the mean work to the probe, and $v$ is the mean rotation rate of the dimeric probe. This model, which was empirically chosen, successfully fits the simulation curves [Fig. S5a, b]. Using the obtained parameters $p$, $q$, $b$, $c$, and $d$, we fitted the experimental data by eye with Eqs. (S18a) and (S18b) and obtained $s = 1000\,\text{Hz}/\mu\text{M}$ [Fig. 3a].

Figure S5c shows the simulated work as a function of rotation rate, for several spring constants $k$. Work does not significantly depend on $k$ in the range of rotation rates we investigated.

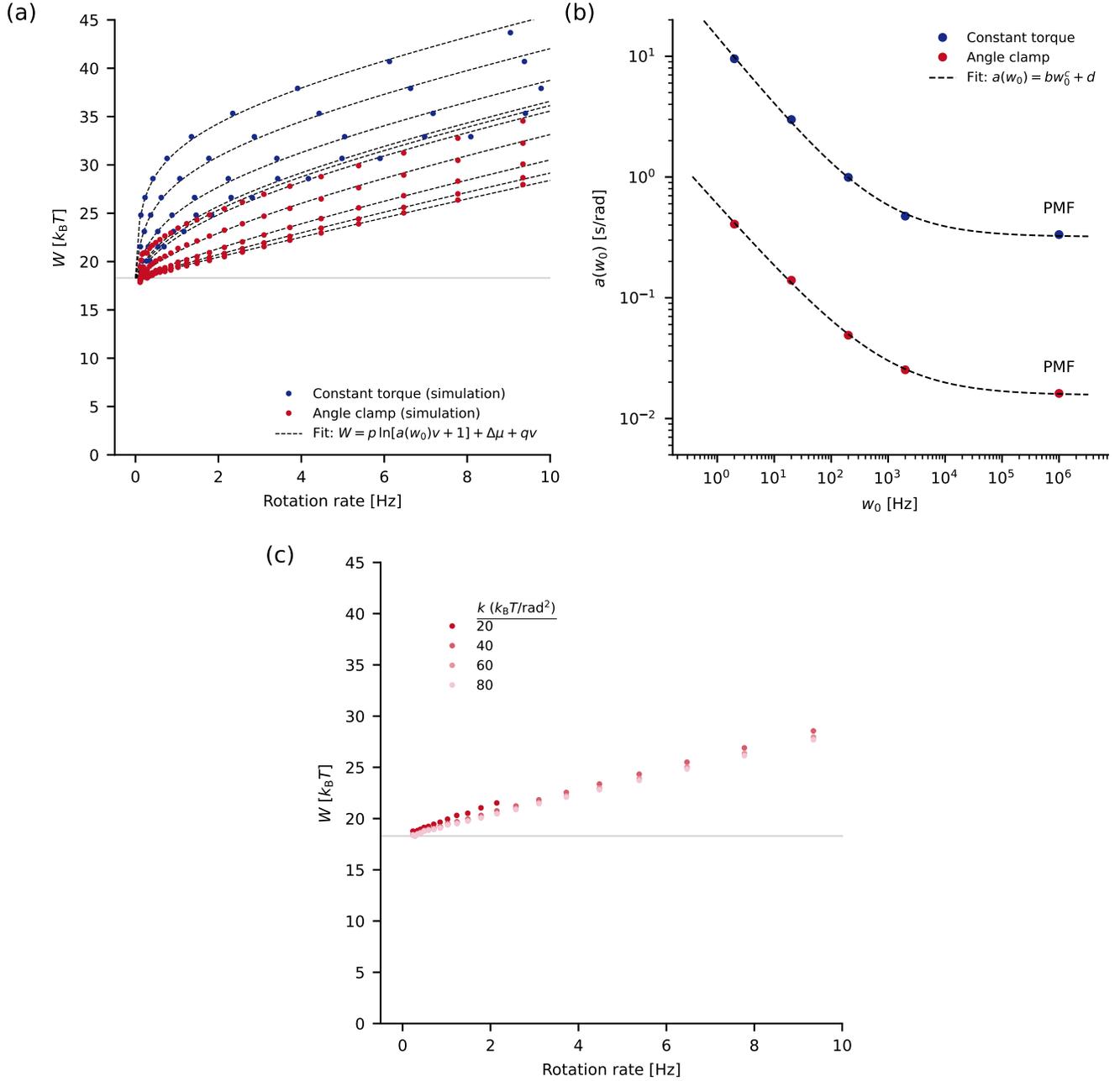

FIG. S5. Simulation. (a) Work per 120° in simulation, as a function of the probe's mean rotation rate $v$, for $w_0 = 2, 20, 200, 2000$ Hz, and $\infty$ from top to bottom ($w_0 = \infty$ was simulated using the PMF, Sec. S2 S2.2). Blue symbols: constant torque; red symbols: angle clamp with spring constant $60\,k_B T/\mathrm{rad}^2$. Dashed curves are fits to $W(v)$ given by Eq. (S18a) with the $w_0$-independent fitting parameters $p = 3\,k_B T$ and $q = 0.13\,k_B T\mathrm{s/rad}$ and $w_0$-dependent fitting parameter $a(w_0)$. (b) Fitting parameter $a(w_0)$ obtained in (a) as a function of $w_0$. Dashed curves are fits to Eq. (S18b) (in the logarithm space of $a$) with the fitting parameters $b = 14$, $c = -0.58$, and $d = 0.32$ for constant torque, and $b = 0.59$, $c = -0.53$, and $d = 0.016$ for angle clamp when $a(w_0)$ and $w_0$ are evaluated in the units of s/rad and Hz, respectively. In fits and the figure, we substitute $w_0 = 10^6$ Hz (a sufficiently large value) for $w_0 = \infty$. (c) Work per 120° in angle-clamp simulations on the PMF ($w_0 \to \infty$) for different trap strengths $k$. In the angle-clamp simulation in (a) and (c), to match experiments we discarded trajectories where the probe escaped from the local minimum of the sinusoidal trapping potential. Since the probe readily escapes from the trap for weak trap strength and fast rotation, for $k = 2\,k_B T/\mathrm{rad}^2$ data is only available with $v \leq 2$ Hz. The dependence of $W$ on $k$ is small. Solid horizontal gray lines in (a) and (c): the reversible work, $W = \Delta\mu$ (18.3$k_B T$).


\fi

\end{document}